%% file: main.tex
\documentclass[10pt,journal,compsoc]{IEEEtran}
\usepackage[nocompress]{cite}
\usepackage{amsmath,amssymb}
\usepackage{graphicx}
\usepackage{booktabs}
\usepackage[table]{xcolor}
\definecolor{TableHeaderShade}{HTML}{EAF0F4}
\usepackage{tikz}
\usetikzlibrary{arrows.meta,positioning}
\usepackage[hidelinks]{hyperref}
\hypersetup{pdftitle={Fast Intent-Driven Service Orchestration with Jev for 6G Edge Networks},pdfauthor={Delong Li, Xu Wang, Haochen Gong, Rui Lang, Guangsheng Yu}}
\begin{document}
\title{Fast Intent-Driven Service Orchestration with Jev for 6G Edge Networks}
\author{Delong Li, Xu Wang, Haochen Gong, Rui Lang, Guangsheng Yu\textsuperscript{*}%
\thanks{The authors are with the School of Electrical, Mechanical and Biomedical Engineering, University of Technology Sydney, Sydney, Australia.
\textsuperscript{*}Corresponding author: Guangsheng Yu (e-mail: \href{mailto:Guangsheng.Yu@uts.edu.au}{Guangsheng.Yu@uts.edu.au}).}}
\IEEEtitleabstractindextext{%
\begin{abstract}
Intent-driven services envisioned for sixth-generation (6G) edge networks must translate changing requirements into executable contracts while wireless requests continue to arrive. Interpretation consumes part of the same deadline budget as transmission, queueing, and execution. We evaluate Jev as a decision model for this stage, asking whether faster contract activation improves service timeliness while preserving interpretation quality. Contracts specify permitted execution locations, deadlines, and priorities; a numerical scheduler uses them to allocate edge work. The evaluation combines live Jev, DeepSeek, Gemini, and self-hosted Qwen responses with packet-level New Radio simulation, mobility events, shared edge queues, and a real image-reading service. Cached interpretation with numerical scheduling substantially improves completion over direct model-selected placement. With correct interpretation on all evaluated contracts, Jev reduces median decision latency by 22.4\% against DeepSeek and 61.9\% against Gemini; completion in the modeled update scenarios rises by 3.50 and 8.35 percentage points. A separate comparison with direct-attribute Qwen retains a 53.0\% latency reduction and a 4.78-point completion gain. Exchanging recorded model and scheduling waits reproduces the comparator completion summaries for all 12 trajectories in each of two modeled comparisons. In the real image service, Jev retains faster decisions and completes 459 requests correctly and on time, compared with 463 for DeepSeek and 435 for Qwen out of 1,080 each. These findings connect fast intent decisions to timely edge service execution and support using Jev at contract activation, with numerical scheduling adapting placements to current network and computing state.
\end{abstract}
}
\maketitle
\IEEEdisplaynontitleabstractindextext
\IEEEpeerreviewmaketitle
\input{sections/introduction}
\input{sections/related-work}
\input{sections/system}
\input{sections/study}
\input{sections/results}
\input{sections/discussion}
\input{sections/conclusion}
\bibliographystyle{IEEEtran}
\bibliography{references}
\appendices
\input{sections/appendix}
\end{document}

%% file: sections/introduction.tex
\section{Introduction}
\IEEEPARstart{S}{ixth-generation} (6G) networks are envisioned to integrate communication and computation for intelligent services~\cite{letaief2019roadmap}. The International Mobile Telecommunications framework for 2030 and beyond (IMT-2030) identifies artificial intelligence (AI) and communication as a usage scenario and anticipates computing services within the network~\cite{itu2023imt}. Mobile edge computing brings services near wireless users, whose quality of service (QoS) depends on both data delivery and computing resources~\cite{shi2016edge,mao2017survey}. Intent-based orchestration expresses the desired outcomes independently of their implementation~\cite{clemm2022intentbased}.

Consider a mobile image-reading service. Mobility changes the serving cell, while competing workloads change edge queues. Separately, a new instruction may restrict processing to one location, withdraw a restriction, or tighten a deadline. The orchestrator must interpret it while wireless requests continue arriving. Delivered images governed by the new requirements wait for their contract to become available. Radio delivery and interpretation waiting thus consume the same response budget. The orchestration problem is to activate the correct requirements early enough for affected requests to finish on time.

Large language models (LLMs) and smaller language models translate intents into network configurations and service specifications~\cite{jacobs2021hey,angi2025llnet,wang2024netconfeval}, often alongside optimization, validation, or agent coordination~\cite{miyaoka2025chatdriven,martins2026intentdriven,parra-ullauri2026rolebased}. Mobile edge operation presents two distinct sources of change: network and computing state evolves during execution, while a new intent changes the requirements that execution must satisfy. A model can select requirements and placements together, or interpret requirements once and let numerical scheduling track evolving resources. Comparing these choices requires accounting for control waiting alongside wireless delivery and edge queues.

Interpretation produces a bounded contract: permitted execution locations, a deadline, and priority. These attributes can be selected from a catalog or returned directly. Jev exposes bounded decisions through a structured application programming interface (API)~\cite{typesafe2026systemone}, making it a candidate for this stage. Three research questions (RQs) guide our evaluation. \emph{RQ1:} How does direct model-selected placement compare with cached interpretation and numerical scheduling? \emph{RQ2:} At matched interpretation quality, how fast is Jev, and can control waiting explain service-completion differences? \emph{RQ3:} How do output interfaces and actual application execution affect the comparison?

We compare live Jev, DeepSeek, Gemini, and self-hosted Qwen responses using packet-level New Radio (NR) traces and shared edge queues. Wireless arrivals, handovers, and control waits share a timeline. Integration comparisons address RQ1; matched-quality comparisons and waiting-time replays address RQ2. For RQ3, we compare two Qwen interfaces and execute a real optical character recognition (OCR) service after simulated NR delivery. Returned text, contract compliance, and deadlines connect interpretation to mobile service outcomes.

\begin{samepage}
The paper makes three contributions:
\begin{enumerate}
\item \textbf{Integration benefit.} We show that cached intent interpretation with numerical scheduling improves mobile edge completion over direct model-selected placement. Jev's updated-contract completion rises from 43.24\% to 95.59\%; DeepSeek and Gemini also benefit.
\item \textbf{Timing attribution.} At matched interpretation correctness, Jev reduces median decision latency by 22.4\% against DeepSeek and 61.9\% against Gemini and improves modeled completion. Replays reproduce the comparator completion summaries by exchanging recorded model and scheduling waits in both matched-quality comparisons.
\item \textbf{Interface and application evidence.} Direct-attribute output brings Qwen to Jev's observed contract correctness, while Jev retains a 53.0\% latency reduction. In the real image service with simulated NR access, Jev remains faster, with completion close to DeepSeek and higher than Qwen.
\end{enumerate}
\end{samepage}

%% file: sections/related-work.tex
\section{Related Work}
\subsection{Intent Translation and Network Orchestration}
Intent-based networking distinguishes desired outcomes from the mechanisms used to realize them~\cite{clemm2022intentbased}. Language interfaces implement this translation at several levels: Lumi addresses network management~\cite{jacobs2021hey}, and Manias et al. study intent extraction for fifth-generation (5G) core networks~\cite{manias2024intentbased}. LLNet uses a small language model to instruct softwarized devices~\cite{angi2025llnet}, while SLM\_netconfig fine-tunes small models for configuration generation~\cite{lira2025network}. NetConfEval evaluates network-configuration tasks~\cite{wang2024netconfeval}, and NetLLM adapts language models to networking tasks~\cite{wu2024netllm}. Together, these studies motivate evaluating both the meaning of a model's output and the cost of obtaining it.

The closest architectural precedent is chat-driven virtual network management, which separates an Interpreter from an optimization-based placement and routing stage~\cite{miyaoka2025chatdriven}. Its comparison of a lightweight classifier with an LLM also makes the quality--latency trade-off explicit. Intent Engine constructs grounded service specifications and assesses translation overhead and downstream placement effects~\cite{islam2026intentpreprint}. Its stated deployment uses human-triggered or low-frequency updates. Our focus is the transition interval itself: newly governed requests continue arriving, and their remaining deadline budgets shrink while a replacement contract is interpreted.

For 6G service orchestration, distributed management introduces a further coordination problem. DMO-GPT uses LLMs for natural-language interaction with operations support systems across operators~\cite{mekrache2026dmogpt}. Martins et al. combine catalog grounding, structural validation, and service decomposition~\cite{martins2026intentdriven}. Brodimas et al. use agentic AI for infrastructure and service orchestration~\cite{brodimas2025intentbased}, and Parra-Ullauri et al. organize agents by business, service, and infrastructure roles~\cite{parra-ullauri2026rolebased}. Our comparison complements these designs by measuring waiting at the contract-interpretation stage under a common execution policy.

Application routing also depends on the relationship between intent and current resources. JAUNT aligns user intent with network state for quality-of-experience-aware tool routing~\cite{li2025jaunt}. From Prompt to Service places a small language model at an edge gateway, where a service registry validates its routing decision before invoking a backend~\cite{nisiotis2026prompt}. These systems place bounded semantic decisions within a larger service. Our measurements follow the requests affected by a contract update and attribute their completion differences through recorded control waits.

\subsection{Efficient Decisions, Interfaces, and Reuse}
Specialized extraction and classification offer several ways to obtain bounded decisions. SetFit fine-tunes sentence representations and a classification head from few labeled examples~\cite{tunstall2022efficient}. GLiNER uses a bidirectional encoder for parallel entity extraction~\cite{zaratiana2024gliner}, and GLiNER2 supports schema-driven structured extraction~\cite{zaratiana2025gliner2}. These alternatives give concrete context to the small-model perspective on repeated agent tasks~\cite{belcak2025small}. Jev provides a decision-oriented interface with bounded choices~\cite{typesafe2026systemone}; our measurements evaluate its role as a contract interpreter.

Generative output also depends on the chosen interface and execution system. JSONSchemaBench evaluates the efficiency, constraint coverage, and quality of structured-output methods~\cite{geng2025jsonschemabench}. SGLang combines language-program execution with cache reuse and structured-decoding optimizations~\cite{zheng2023sglang}, while vLLM improves serving through PagedAttention memory management~\cite{kwon2023efficient}. Our local comparison holds the serving configuration fixed and examines an application-level choice: selecting a catalog identifier versus specifying its semantic attributes. The latter produces a longer answer but substantially improves correctness in this workload, changing the appropriate baseline for the latency comparison.

Model routing and cascades address a complementary selection problem. FrugalGPT selects combinations of LLM calls to trade cost against answer quality~\cite{chen2023frugalgptb}; RouteLLM learns to choose between stronger and weaker models using preference data~\cite{ong2024routellma}. Our study compares interpreters at a fixed orchestration stage, where decision waiting consumes the application's deadline budget. Interpreter selection can therefore be assessed through timely completion alongside answer quality and API fees.

GPTCache demonstrates the value of reusing language-model responses~\cite{bang2023gptcache}. In our system, reuse is exact and version-aware: it avoids interpreting unchanged contract text, while scheduling continues to consume current numerical observations. The integration comparison measures the combined service effect of contract reuse and numerical scheduling.

\subsection{Communication and Computing for 6G Services}
The 6G vision connects AI-enabled network operation with the provision of computing services across devices and network infrastructure~\cite{letaief2019roadmap,itu2023imt}. Communication, placement, and computation jointly determine mobile edge response time~\cite{mao2017survey}. Research on low-latency delivery across the edge--cloud continuum treats service performance over its execution lifetime~\cite{santos2021lowlatency}. For prediction pipelines, InferLine combines provisioning and scaling to meet end-to-end latency objectives under changing arrivals~\cite{crankshaw2020inferline}. Our service-level evaluation brings intent-activation waiting into this timing analysis: wireless delivery consumes part of a request's budget before the request can use its interpreted requirements and edge resources.

We use the 5G-LENA NR simulator~\cite{patriciello2019nr} to provide packet delivery and mobility events, then incorporate measured model waits into shared service queues. Koutlia et al. document the simulator's calibration against Third Generation Partnership Project (3GPP) reference scenarios~\cite{koutlia2022calibration}. In the application experiment, the queue workers execute OCR on IIIT5K images~\cite{mishra2012scene} using Tesseract~\cite{smith2007overview}. The application outputs let us distinguish a timely modeled completion from a timely and correctly recognized image.

%% file: sections/system.tex
\section{Intent-Driven Edge Orchestration}
\label{sec:system}
\subsection{Wireless Access and Edge Resources}
The service path consists of mobile users, wireless access, an orchestrator, and edge workers. In the trace-driven model, handover changes a user's serving cell and hence which worker has the shorter transport path. Shared queues, available computing capacity, and the overhead of changing placement also affect the preferred worker. Service requirements constrain these placement choices. We place the interpreter at this service-orchestration layer: it supplies the requirements under which the numerical scheduler allocates edge work. Wireless delivery and mobility events continue while either stage is running.

\subsection{Contracts and Request Versions}
Let $u$ index a user and $v$ a contract version shared across the users. Versions are numbered in order of issue, with issue time $\tau_v$; an initial version exists at or before the first request. For each user, $\mathcal{A}_{u,v}$ is the permitted set of edge workers, $d_{u,v}>0$ the send-to-completion budget, and $p_{u,v}>0$ the priority, with larger values served first. These requirements form the intended service contract
\begin{equation}
c_{u,v}=(\mathcal{A}_{u,v},d_{u,v},p_{u,v}).
\label{eq:contract}
\end{equation}
The two workers are denoted A and B, giving three supported location sets: A only, B only, and either worker. A natural-language description expresses each contract.

For request $i$, let $u_i$ be its user and $s_i$ its send time. Its version $v_i$ is fixed at generation:
\begin{equation}
v_i=\max\{v:\tau_v\le s_i\}.
\end{equation}
Scoring uses $c_{u_i,v_i}$ even if another version is issued before the request finishes. Consequently, a later update does not retrospectively change the requirements of work already submitted. The evaluator stores canonical contracts separately from the text passed to a model. The model returns an interpreted contract $\widehat c_{u,v}$ with attributes $\widehat{\mathcal A}_{u,v}$, $\widehat d_{u,v}$, and $\widehat p_{u,v}$. A shared validator checks format and supported values. Validation permits execution; semantic correctness is scored separately against the canonical contract.

\subsection{Semantic Interpretation and Placement}
We compare two integration choices. In direct orchestration, the model receives the current contract descriptions and observed numerical state, then selects both contract profiles and the joint placement of the users. The shared execution layer checks the returned plan against the interpreted location permissions before applying it. Numerical state includes current queues, available service capacity, communication delay estimates, and existing placements.

In cached interpretation, the model only translates the current contract descriptions. The cache identity includes the descriptions, version, and catalog identity. A change to these inputs requires fresh interpretation. A handover or capacity event can reuse the existing contract while triggering placement reconsideration. A numerical scheduler chooses from the permitted combinations using the state available after interpretation finishes. This separates changes in service requirements from changes in the network and computing resources used to satisfy them.

The trace-driven scheduler enumerates the permitted joint placements and scores them using estimated queueing and processing delay, current offered load, communication delay, and placement-change overhead. Delay is normalized by each interpreted deadline and weighted by its interpreted priority. This is a shared heuristic used to compare interpreters. The real image service uses the same separation of responsibilities, with per-request placement at the allowed worker having the earliest predicted finish. Each worker serves queued work in non-preemptive priority order, breaking priority ties by deadline. The precise scoring rule is given in the appendix.

The structured-input reference receives canonical contracts directly and uses the same numerical scheduler. It measures service execution when an upstream system already supplies structured requirements and no language interpretation is needed.

\begin{figure}[t]
\centering
\begin{tikzpicture}[font=\footnotesize,>=Latex,
 box/.style={draw=black!55,rounded corners=2pt,align=center,minimum height=8mm,text width=24mm},
 arr/.style={->,line width=.65pt}]
\node[box] (text) at (0,0) {Versioned\newline intent text};
\node[box,fill=teal!7] (parse) at (3,0) {Interpretation\newline and validation};
\node[box] (cache) at (6,0) {Contract\newline cache};
\node[box] (radio) at (0,-1.65) {Mobile request\newline arrivals};
\node[box,fill=teal!7] (gate) at (3,-1.65) {Version gate\newline and scheduler};
\node[box] (edge) at (6,-1.65) {Edge queues\newline and execution};
\draw[arr] (text)--(parse);
\draw[arr] (parse)--(cache);
\draw[arr] (radio)--(gate);
\draw[arr] (cache.south)--(gate.north east);
\draw[arr] (gate)--(edge);
\draw[arr] (edge.south)--++(0,-.6)-|node[pos=.25,below,font=\scriptsize]{current numerical state}(gate.south);
\draw[dashed,black!45] (-1.35,-.85)--(7.35,-.85);
\end{tikzpicture}
\caption{Intent-driven orchestration above wireless access. Semantic interpretation activates a service contract at first use and updates. Radio-delivered requests then pass the version gate to edge execution. Numerical scheduling uses current network and computing state, while unchanged contracts are reused.}
\label{fig:system}
\end{figure}
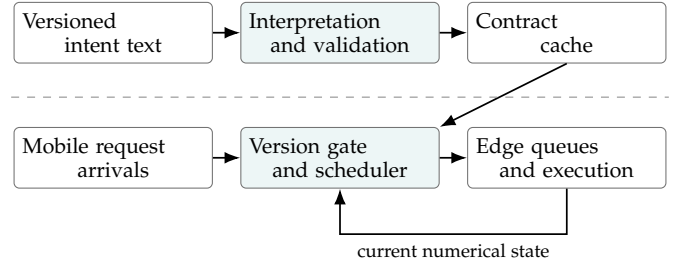

\subsection{Wireless Delivery and the Service Deadline}
Figure~\ref{fig:system} locates the semantic stage relative to arriving requests. Let $a_v$ be the time at which a validated interpretation of version $v$ is installed, and $b_i$ the radio-delivery time of request $i$. For a delivered request whose contract becomes available, define the earliest time permitted by version readiness, $r_i$, and the associated gate wait, $g_i$, as
\begin{equation}
r_i=\max(b_i,a_{v_i}),\qquad
g_i=\max(0,a_{v_i}-b_i).
\label{eq:gate}
\end{equation}
Placement changes and service queues can delay execution beyond $r_i$. Radio loss or failure to activate a usable contract prevents successful completion; Eq.~\eqref{eq:gate} applies to requests that pass those prerequisites.

The remaining budget at version readiness makes the communication--orchestration coupling explicit:
\begin{equation}
s_i+d_{u_i,v_i}-r_i
=d_{u_i,v_i}-(b_i-s_i)-g_i.
\label{eq:remaining-budget}
\end{equation}
Here $b_i-s_i$ is elapsed wireless-delivery time, and $g_i$ is any additional waiting for the contract after delivery. A negative value means that these two intervals have already exhausted the deadline. The remaining time must accommodate scheduling, transport to the selected worker, queueing, and execution. Earlier activation reduces $g_i$ for requests waiting on that version and leaves more time for these stages. One interpretation can therefore benefit many requests; subsequent requests reuse the installed contract.

Decision latency is measured at the client from invocation until the response is available to the caller. It includes transport, provider processing, and queueing along that path. Scheduling duration is recorded separately; together, the two measurements define the control waits used in our replay analysis. During every measured wait, the trace-driven model advances arrivals, radio events, queues, and capacity changes. The real service uses asynchronous requests on a common wall-clock timeline. Neither execution pauses its workload clock to await the model.

\subsection{Quality and Completion Metrics}
Semantic correctness requires exact agreement of all three interpreted contract attributes with their canonical values. This is evaluated separately from the service outcome.
Let $f_i$ be request completion time and $e_i$ its execution worker; for the image service, completion means that the result has returned to the orchestrator. Write $\mathbf{1}\{E\}$ for the indicator of condition $E$. In the modeled service, define
\begin{equation}
y_i^{\mathrm{model}}=
\mathbf{1}\{f_i\le s_i+d_{u_i,v_i}\}
\mathbf{1}\{e_i\in\mathcal A_{u_i,v_i}\},
\label{eq:success}
\end{equation}
with the indicator set to zero for unfinished requests. Priority affects queue order but is not a separate success predicate in this modeled metric. For the image service, let $z_i$ equal one when a successfully returned text matches the normalized reference, and zero otherwise. Normalization preserves case, applies Unicode Normalization Form C, and trims boundary whitespace. Let $\widehat p_i$ be the interpreted priority used for request $i$. Using the same deadline and location checks, its success indicator is
\begin{equation}
y_i^{\mathrm{image}}=y_i^{\mathrm{model}}z_i
\mathbf{1}\{\widehat p_i=p_{u_i,v_i}\}.
\label{eq:image-success}
\end{equation}
Both indicators are zero for unfinished requests.

For a trajectory containing $N>0$ submitted foreground requests, the completion fraction is $N^{-1}\sum_{i=1}^{N}y_i$, using the indicator appropriate to that experiment. Radio losses and unsuccessful requests remain in the denominator. We report modeled and image-service outcomes separately because the latter also requires a correct application output.

%% file: sections/study.tex
\section{Mobile Edge Evaluation Design}
\label{sec:study}
The four measurement blocks connect integration choice, quality-matched timing, interface sensitivity, and application outcomes. Table~\ref{tab:design} summarizes their scale. Jev, DeepSeek, and Gemini are accessed through OpenRouter; Qwen is self-hosted. Generative comparators return compact structured responses. Model versions, output settings, radio parameters, and implementation details are in the appendix.

The common network setting combines mobile access, handovers between cells, competing edge workloads, and changing service contracts. Contract updates are specified independently of mobility events. Each comparison uses the same NR delivery and handover traces, so an interpreter's waiting time changes the service state encountered after its response without changing the radio input. Edge actions do not feed back into the radio simulation. This isolates the service-orchestration stage within the communication and computation path.

\begin{table}[t]
\centering
\caption{Study size by measurement block. Requests are counted per trajectory.}
\label{tab:design}
\footnotesize\setlength{\tabcolsep}{6pt}\renewcommand{\arraystretch}{1.15}
\begin{tabular}{@{}lrr@{}}\toprule
\rowcolor{TableHeaderShade}
Block & Trajectories & Requests\\\midrule
Integration & 14 & 22,500\\
Cloud & 60 & 22,500\\
Local interface & 36 & 22,500\\
Image service & 16 & 270\\\bottomrule
\end{tabular}
\end{table}

The integration block compares direct placement with cached interpretation and numerical scheduling under stable and updated contracts. It includes Jev, DeepSeek, Gemini, and the structured-input reference; both responsibility assignment and contract reuse change between the two model configurations. The subsequent cloud and local blocks use cached interpretation with a common scheduler. Their four scenarios are a stable contract, withdrawal of a location restriction, changed location permissions, and reversal of priorities. Each update scenario also specifies its current deadline values. The transitions differ in contract content and update time. The cloud block uses 21 distinct contract descriptions, each scored in three timing repeats. Each invocation interprets a batch of three user contracts, giving 21 calls and 63 contract checks per model. The local block reuses that text set and measures a new Jev reference alongside the two Qwen interfaces.

The cloud block includes Jev, DeepSeek, Gemini, profile-output Qwen, and the structured-input reference. The principal comparison uses the three cloud models, which interpret all evaluated contracts correctly. The local block compares Jev with two interfaces to the same Qwen deployment: profile selection and direct attributes. The latter matches Jev's observed correctness and becomes the principal local comparator. This baseline selection follows the exploratory measurements; both Qwen interfaces remain in the reported results.

The image block tests the integration on 96 public scene-text images. Twelve contract descriptions govern three users over four versions. Each trajectory submits 270 image requests, and each method executes four trajectories, for 1,080 requests per method. Two mobility and background-load traces are shared across methods. All radio fragments of an image must arrive before it is released to the real service. Model invocation, transfer, and OCR then share a wall-clock timeline; the deadline starts at the image's original send time. We check the full trajectory and the fixed 1.5-second windows following contract updates.

Conditions are interleaved using a fixed randomized order. We report all timing repeats and retain long responses. Means, quantiles, and observed ranges are descriptive summaries of these timing measurements. Requests within a trajectory share contracts, queues, and radio conditions, so their count is not a count of independent experimental replicates. Response latency uses all interpretation calls in its block, whereas the updated-scenario completion aggregate excludes the stable scenario. Within each block, a relative latency reduction is one minus Jev's median divided by the comparator's median. Completion differences subtract the comparator's mean percentage from Jev's; they are reported in percentage points. Exact configuration and validation details are collected in the appendix.

%% file: sections/results.tex
\section{Results}
\subsection{Model Integration and Completion}
\label{sec:integration}
\begin{figure}[t]
\centering\includegraphics[width=\columnwidth]{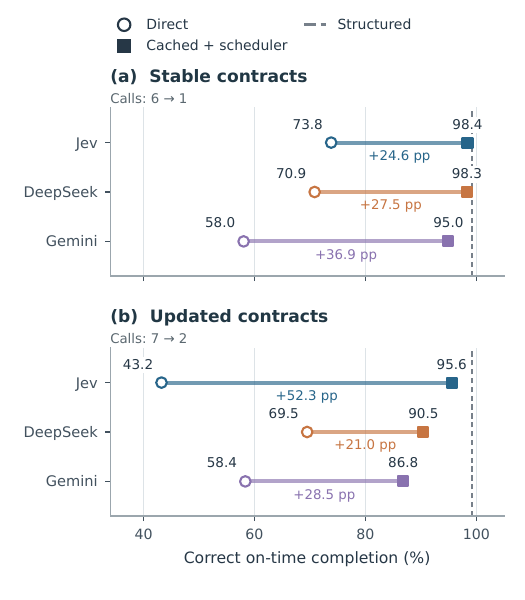}
\caption{Integration comparison with modeled edge execution. Endpoint labels give completion percentages for direct placement (open circles) and cached interpretation with numerical scheduling (squares); connector labels give percentage-point gains. Dashed lines show the structured-input reference. Call counts compare direct and cached configurations. Each mark is one trajectory with 22,500 requests; scheduling responsibility and semantic reuse change together.}
\label{fig:integration}
\end{figure}

Figure~\ref{fig:integration} compares direct model-selected placement with cached interpretation and numerical scheduling. Under updated contracts, the respective completion counts are 9,729 versus 21,507 for Jev, 15,639 versus 20,355 for DeepSeek, and 13,130 versus 19,537 for Gemini, each out of 22,500 requests. The structured-input reference completes 22,312. All three models interpret the contract profiles correctly in this block. The performance difference therefore involves how those contracts are converted into a placement and when the placement takes effect.

The numerical scheduler jointly accounts for offered load, worker capacity, and transport delay relative to the serving cell. A placement can obey all location permissions yet concentrate enough work at one worker to delay many requests. Separating these resource decisions from semantic interpretation substantially improves completion for each tested model. For Jev, the updated-condition completion fraction increases from 43.24\% to 95.59\%.

Reuse also reduces interpretation frequency. Stable contracts require six model calls in direct orchestration and one in the cached configuration. Updated contracts require seven and two, respectively. Each cached trajectory has five hits, while the scheduler continues to reconsider placement using current observations. We use this integration in the remaining comparisons: interpret new contracts once and schedule against current numerical state.

\subsection{Correct Interpretation with Lower Response Time}
\label{sec:latency}
\begin{figure*}[t]
\centering\includegraphics[width=\textwidth]{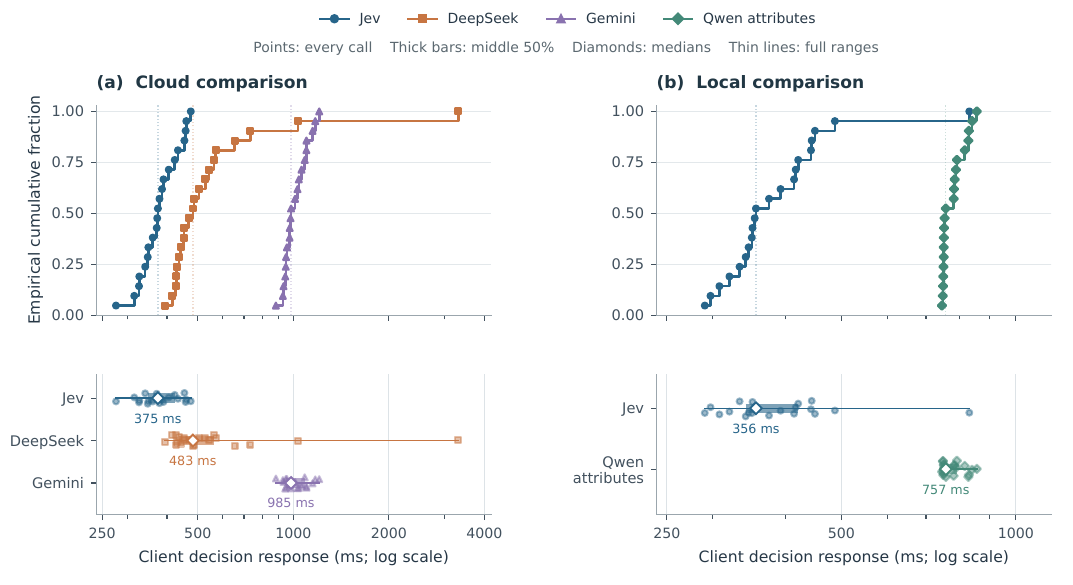}
\caption{Measured decision-response distributions. Upper plots show empirical cumulative fractions for all 21 calls per model. Lower strips retain every call, with deterministic vertical offsets for visibility; thick bars span the 25th--75th percentiles, diamonds and labels give medians, and thin lines show full ranges. The cloud and local blocks have separate contemporaneous Jev measurements. Logarithmic axes retain the complete ranges, including DeepSeek's 3.315-second response. Quantiles describe calls, not confidence intervals.}
\label{fig:latencies}
\end{figure*}
\input{tables/cloud-results}
In the cloud block, Jev, DeepSeek, and Gemini each correctly interpret all 63 evaluated contracts, corresponding to 21 distinct descriptions repeated three times. Table~\ref{tab:cloud} reports their response medians and updated-scenario completion. Jev's median is 374.6 ms, compared with 482.9 ms for DeepSeek and 984.5 ms for Gemini, reductions of 22.4\% and 61.9\%. Pairing calls by scenario, repeat, and version, Jev responds faster in 20 of 21 comparisons with DeepSeek and all 21 comparisons with Gemini.

Figure~\ref{fig:latencies} shows the complete latency distributions. The cloud result includes a 3.315-second DeepSeek call, making its tail appreciably longer than its median. That call remains in the distribution and delays activation in its service trajectory, allowing the completion analysis to reflect the observed long tail.

The separate local block uses a contemporaneous Jev measurement and Qwen's direct-attribute interface. Both interpret all 63 contracts correctly. Their medians are 356.2 and 757.1 ms, giving Jev a 53.0\% reduction. The comparison measures the response available to the orchestrator across each deployment path, which contributes to activation waiting in Eq.~\eqref{eq:gate}.

\subsection{Contract Updates Expose the Timing Difference}
\label{sec:completion}
\begin{figure*}[t]
\centering\includegraphics[width=\textwidth]{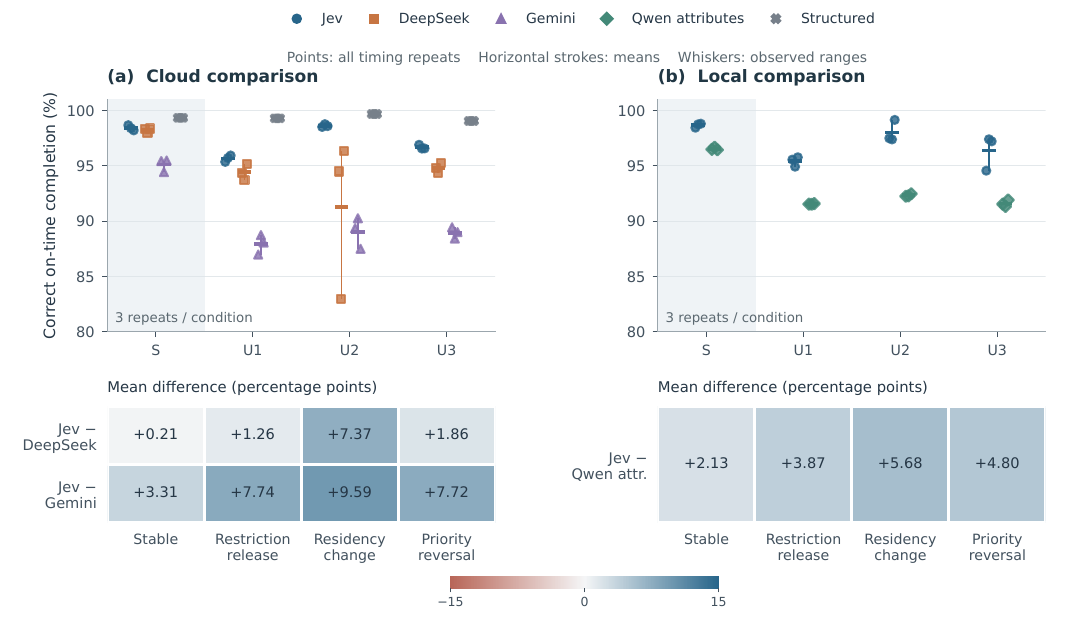}
\caption{Correct on-time completion in modeled edge execution. Upper plots show every timing repeat, means (horizontal strokes), and observed ranges (vertical segments). S is stable; U1--U3 are the three update scenarios named below the matrices. Lower cells show Jev-minus-comparator mean differences in percentage points on a common color scale. Each trajectory has 22,500 requests. The local block uses direct-attribute Qwen and its own Jev measurements.}
\label{fig:completion}
\end{figure*}

Across the three update scenarios in the cloud block, mean completion is 97.00\% for Jev, 93.51\% for DeepSeek, and 88.65\% for Gemini. Jev's differences are 3.50 and 8.35 percentage points. Figure~\ref{fig:completion} shows that the size of the difference depends on the transition. Restriction withdrawal, permission replacement, and priority changes alter which placements are useful and which requests have the least remaining slack.

The stable cloud scenario has a much smaller Jev--DeepSeek difference: 98.44\% versus 98.23\%. Both still incur first-use interpretation; subsequent telemetry events reuse the contract. Stable here describes the service requirements: mobility and resource events still occur. The shared scheduler handles that evolving state using the cached contract. The larger updated-condition differences arise when renewed interpretation consumes additional time from the service budget in Eq.~\eqref{eq:remaining-budget}.

The local block shows the same direction with a quality-matched comparator. Mean updated-scenario completion is 96.62\% for Jev and 91.84\% for direct-attribute Qwen, a 4.78-point difference. Stable-condition completion is 98.68\% and 96.55\%, respectively.

\subsection{Completion under Exchanged Waiting Times}
\label{sec:intervention}
To test the timing explanation, we replay each trajectory with the original interpreted contracts and common execution model, exchanging the recorded model waits and numerical scheduling times. The radio events, workload, service rules, and scoring contracts remain fixed. This post-measurement analysis introduces no new model calls.

Let $F(X,C,W)$ denote the completion summary produced by the deterministic execution model. Here $X$ contains the shared radio and workload input, $C$ the sequence of interpreted contracts, and $W$ the recorded model and scheduling durations. For Jev and a comparator, the intervention evaluates $F(X,C_{\mathrm{Jev}},W_{\mathrm{comp}})$ against the comparator's observed modeled summary $F(X,C_{\mathrm{comp}},W_{\mathrm{comp}})$. For the matched-quality pairs, the contract sequences agree. Substituting the comparator's waits tests whether those recorded delays reproduce its service result under the common scheduler.

Each replay set covers four scenarios and three timing repeats; a match requires the full modeled summary to agree. All 12 cloud trajectories exactly reproduce the DeepSeek summaries after this substitution. All 12 local trajectories likewise reproduce direct-attribute Qwen. Thus, the recorded control waits reproduce the modeled completion differences with contract semantics, radio inputs, and scheduling policy held fixed.

\subsection{A Better Local Interface Changes the Baseline}
\label{sec:interface}
\input{tables/interface-results}
The initial Qwen interface selects identifiers from an 18-profile catalog. It correctly interprets 33 of 63 evaluated contracts. We then change the interface to return location permissions, deadline, and priority directly. The weights, serving instance, business inputs, and non-thinking setting are shared between these two local conditions.

Table~\ref{tab:interface} shows the joint change in quality and time. Direct attribute output reaches 63/63 correct interpretations. Median generated length rises from 28 to 56 tokens, and median response increases from 447.9 to 757.1 ms. Despite the additional waiting, mean updated-scenario completion improves from 85.17\% to 91.84\%. In the restriction-release scenario, the profile interface also produces 2,750 location violations in each timing repeat; these disappear with direct attributes.

The longer attribute response yields better service completion by improving interpretation. It supplies semantic values directly instead of mapping them to catalog identifiers. This interface change combines a different prompt, schema, and output length. Direct-attribute Qwen is consequently the principal local baseline in Sections~\ref{sec:latency}--\ref{sec:intervention} and the application study.

\subsection{Mobile Image Service over NR Access}
\label{sec:application}
\input{tables/application-results}
The image service preserves the observed semantic quality: Jev, DeepSeek, and direct-attribute Qwen each correctly interpret 48 of 48 contracts across four runs. Their median decision responses are 408.9, 497.0, and 759.9 ms. Jev therefore retains a 17.7\% reduction against DeepSeek and a 46.2\% reduction against Qwen in this application setting.

Table~\ref{tab:application} reports actual correct, on-time image results. Jev completes 459 of 1,080 requests, DeepSeek 463, Qwen 435, and the structured-input reference 473. Jev's full-trajectory result is within four requests, or 0.37 percentage points, of DeepSeek and exceeds Qwen by 24 requests, or 2.22 points. The application therefore retains faster interpretation with similar observed completion to DeepSeek and higher completion than Qwen.

\begin{figure*}[t]
\centering\includegraphics[width=\textwidth]{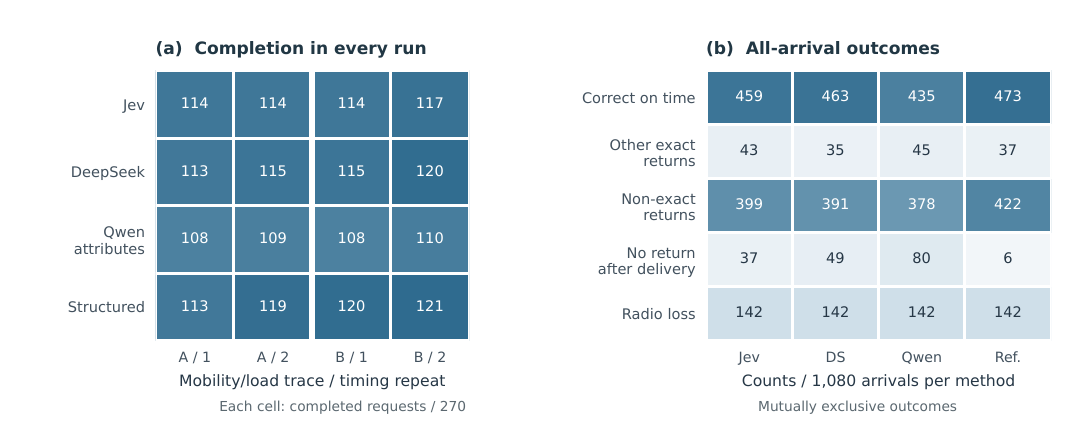}
\caption{Real image-service execution. (a) Correct on-time counts in all 16 runs, each with 270 arrivals; A and B are the two mobility/load traces, each repeated twice. (b) Mutually exclusive outcomes over all 1,080 arrivals per method. Other exact returns have correct text but fail the full completion criterion; non-exact returns have an incorrect text result. No return after delivery excludes radio losses. DS, Qwen, and Ref. denote DeepSeek, direct-attribute Qwen, and structured input. Color scales run from zero to 125 in (a) and zero to 500 in (b); cell labels give exact counts.}
\label{fig:application}
\end{figure*}

Figure~\ref{fig:application} exposes the per-run comparisons. Jev's differences from DeepSeek are $+1$, $-1$, $-1$, and $-3$ requests; its differences from Qwen are $+6$, $+5$, $+6$, and $+7$. Completion also depends on recognizable text, transfer, and queueing; decision-latency ordering alone does not determine the application outcome.

The fixed update windows contain 180 requests per method. Jev completes 64 correctly and on time, compared with 62 for DeepSeek, 45 for Qwen, and 71 for the structured reference. These windows locate the interval in which fresh interpretation is relevant, while the full-trajectory totals retain the behavior outside updates. The full-run denominator also retains 142 radio-lost images per method. Shared radio delivery makes this loss count identical across interpreters; faster activation benefits delivered requests waiting for a contract. Of the returned results, exact normalized text is obtained for 502 requests with Jev, 498 with DeepSeek, 480 with Qwen, and 510 with the structured reference; the stricter completion counts additionally impose the deadline and contract requirements.

All methods have zero observed location violations in this block. Across the complete matrix, 3,964 foreground and background OCR executions match the service-side records. Every selected image has an actual OCR output, and repeated executions of the same image return consistent text. The completion metric thus counts verified service outputs.

%% file: tables/cloud-results.tex
\begin{table}[t]
\centering
\caption{Cloud comparison. Response cells give the median and [25th, 75th] percentiles in milliseconds over 21 calls per model. Contract accuracy uses 63 checks; completion is the mean over nine updated-contract trajectories.}
\label{tab:cloud}
\footnotesize\setlength{\tabcolsep}{4pt}\renewcommand{\arraystretch}{1.16}
\begin{tabular}{@{}lrrr@{}}\toprule
& \multicolumn{2}{c}{Interpretation} & Service\\
\cmidrule(lr){2-3}\cmidrule(l){4-4}
\rowcolor{TableHeaderShade}
Model & Correct & Response (ms) & Updated (\%)\\\midrule
Jev & 63/63 & \shortstack{374.6\\{\scriptsize [348.1, 423.4]}} & 97.00\\
DeepSeek & 63/63 & \shortstack{482.9\\{\scriptsize [436.8, 562.7]}} & 93.51\\
Gemini & 63/63 & \shortstack{984.5\\{\scriptsize [949.9, 1087.5]}} & 88.65\\
\bottomrule\end{tabular}\end{table}

%% file: tables/interface-results.tex
\begin{table*}[t]
\centering
\caption{Local interface comparison over 21 calls and 63 contract checks per interface. Completion rates average three stable or nine updated trajectories. Tokens are medians; NC denotes non-comparable token accounting.}
\label{tab:interface}
\footnotesize\setlength{\tabcolsep}{5pt}\renewcommand{\arraystretch}{1.15}
\begin{tabular}{@{}llrrrrrr@{}}\toprule
& & \multicolumn{2}{c}{Interpretation} & \multicolumn{2}{c}{Response (ms)} & \multicolumn{2}{c}{Completion (\%)}\\
\cmidrule(lr){3-4}\cmidrule(lr){5-6}\cmidrule(l){7-8}
\rowcolor{TableHeaderShade}
Model & Interface & Correct & Tokens & Median & [25th, 75th] & Stable & Updated\\\midrule
Jev & Choice & 63/63 & NC & 356.2 & [341.9, 421.2] & 98.68 & 96.62\\
Qwen & Profiles & 33/63 & 28 & 447.9 & [422.1, 466.1] & 87.31 & 85.17\\
Qwen & Attributes & 63/63 & 56 & 757.1 & [750.8, 791.2] & 96.55 & 91.84\\
\bottomrule\end{tabular}\end{table*}

%% file: tables/application-results.tex
\begin{table*}[t]
\centering
\caption{Real image service with simulated NR access. Response cells show median [25th, 75th percentiles] over 16 calls per interpreter. Execution columns retain all 1,080 arrivals per method; exact text is a subset of returned results, and correct on-time completion also imposes contract and deadline requirements. Update windows are fixed 1.5 s intervals after the three updates (180 requests per method). API fees are totals; N/A denotes no billed model API.}
\label{tab:application}
\footnotesize\setlength{\tabcolsep}{5pt}\renewcommand{\arraystretch}{1.15}
\begin{tabular}{@{}lrrrrrrr@{}}\toprule
& \multicolumn{2}{c}{Contract interpretation} & \multicolumn{3}{c}{Execution / 1,080 arrivals} & \multicolumn{1}{c}{Updates} & \multicolumn{1}{c}{Cost}\\
\cmidrule(lr){2-3}\cmidrule(lr){4-6}\cmidrule(lr){7-7}\cmidrule(l){8-8}
\rowcolor{TableHeaderShade}
Method & Correct & Response (ms) & Returned & Exact text & Correct on time & Correct / 180 & API (US\$)\\\midrule
Jev & 48/48 & \shortstack{408.9\\{\scriptsize [341.4, 482.9]}} & 901 & 502 & \shortstack{459\\{\scriptsize (42.50\%)}} & \shortstack{64\\{\scriptsize (35.56\%)}} & 0.001616\\
DeepSeek & 48/48 & \shortstack{497.0\\{\scriptsize [435.1, 566.4]}} & 889 & 498 & \shortstack{463\\{\scriptsize (42.87\%)}} & \shortstack{62\\{\scriptsize (34.44\%)}} & 0.000939\\
Qwen attributes & 48/48 & \shortstack{759.9\\{\scriptsize [754.0, 791.1]}} & 858 & 480 & \shortstack{435\\{\scriptsize (40.28\%)}} & \shortstack{45\\{\scriptsize (25.00\%)}} & N/A\\
Structured & Given & No call & 932 & 510 & \shortstack{473\\{\scriptsize (43.80\%)}} & \shortstack{71\\{\scriptsize (39.44\%)}} & N/A\\
\bottomrule\end{tabular}\end{table*}

%% file: sections/discussion.tex
\section{6G Service Orchestration}
\subsection{Preserving the Budget after Wireless Delivery}
For 6G mobile edge services, a short wireless transmission time is only one part of timely execution. A delivered request may still be blocked until its new service requirements become usable. Equation~\eqref{eq:remaining-budget} identifies this activation wait within the same send-to-completion budget. The measured-wait replays show how its timing, together with scheduling duration, translates into modeled completion differences under shared radio inputs and a common scheduler.

This supports treating intent-activation delay as part of the service QoS budget. Its value is clearest during transitions, when one interpretation can release many waiting requests. The image-service results show why both transition-window and full-trajectory completion matter: Jev completes more requests during updates than either generative comparator, while its full-run completion remains close to DeepSeek. These measurements connect interpretation speed to useful results returned after wireless delivery.

\subsection{Separating Intent Changes from Network Dynamics}
Mobility and changing edge load require repeated resource decisions even when the service intent remains unchanged. The integration results support caching the interpreted contract and using current numerical observations for those decisions. Serving-cell changes can alter the transport cost of a placement without changing its permission; a newly restricted location instead requires a new contract. Distinguishing these events avoids repeated language interpretation on the numerical scheduling path and retains adaptation to network and computing state. Fast decision models are most useful at first interpretation and subsequent contract updates.

\subsection{Choosing the Interpreter and Its Interface}
Interpreter selection should consider the output interface together with the model. Qwen's direct attributes improve correctness and completion despite a longer response, making that interface the relevant local comparison for Jev. The result favors explicit contract fields when a compact catalog identifier imposes a difficult mapping task. Evaluating the stronger interface retains Jev's latency advantage while comparing decisions of matched observed quality.

Latency and price favor different hosted models in these measurements. Across the cloud block's 21 calls per model, Jev reports US\$0.002125, DeepSeek US\$0.001262, and Gemini US\$0.009698. The image-service fees in Table~\ref{tab:application} also favor DeepSeek. Jev supplies the faster measured response; DeepSeek has the lower billed interpretation fee. Self-hosted Qwen has no per-call API invoice here, but uses a separate compute allocation. A deployment can weigh these fees against the completion benefit of faster activation.

\subsection{Study Scope and Next Steps}
The latency comparisons include the model, serving system, and transport path. Our replays exchange recorded model and scheduling waits together, measuring their effect on execution without attributing the delay to individual provider components. NR simulation supplies the wireless-access evidence for this study of 6G service orchestration; the image service adds actual transfers, queues, and OCR, with two logical workers on one host.

The experiments use a bounded contract catalog and authored transitions. Timing repeats reuse the same inputs, and the application adds two mobility/load traces; requests sharing a trajectory are dependent observations. Completion differences are descriptive, including the four-request Jev--DeepSeek gap in the image service. Varying mobility, link load, and distributed edge capacity independently would test how much interpretation waiting each communication and computing regime can tolerate. Independently sampled contracts would extend the semantic coverage.

%% file: sections/conclusion.tex
\section{Conclusion}
Fast intent-driven service orchestration for 6G edge networks depends on activating the right requirements in time for wireless requests to use the available computing resources. Our evaluation places Jev at this contract-interpretation stage. In the NR-driven workloads, cached interpretation with numerical scheduling improves completion over direct model-selected placement. Jev reduces decision latency at matched contract correctness against hosted generative models and a stronger local Qwen interface. Exchanging recorded control waits reproduces the modeled completion differences, connecting faster decisions to the budget available for service execution. Direct attribute output improves Qwen's correctness and downstream completion, demonstrating the importance of the comparison interface. The real image service retains Jev's faster interpretation with completion close to DeepSeek and higher than Qwen. Together, these results support using fast decision models to activate changing service contracts while numerical scheduling adapts their execution to network and computing state.

%% file: sections/appendix.tex
\section{Implementation and Reproduction Details}
\label{app:implementation}
\subsection{Models and Interfaces}
Measurements were collected on September 19, 2026. The hosted endpoints are TypeSafe Jev 1.13, DeepSeek V4.1 Flash through Together, and Gemini 3.1 Flash Lite through Google AI Studio, accessed through OpenRouter with provider fallback disabled. Jev uses native Choice questions. DeepSeek and Gemini return strict JavaScript Object Notation (JSON) schemas at temperature zero, with token limits of 128 and 256, respectively. DeepSeek reasoning is disabled; Gemini reasoning is minimal. Responses are checked for identity, termination, supported values, and expected fields.

Qwen3.5-4B runs through vLLM~\cite{kwon2023efficient} version 0.29.0 on one NVIDIA L40 graphics processing unit (GPU), using 16-bit brain floating-point precision, a 32,768-token context limit, and at most four sequences. Thinking is disabled. Its two interfaces share weights and serving configuration. Profile output selects among 18 combinations of location, deadline, and priority; direct attributes are normalized to that representation without evaluation labels. The direct interface has a 256-token limit. Client timing includes the access path; server time is recorded separately. API fees use returned usage records.

\subsection{Radio and Modeled Service}
The implementation uses ns-3.48 and 5G-LENA v5.1~\cite{patriciello2019nr}, seed 42 and run 1. Two base stations are 500 m apart. Three user equipment (UE) instances include two moving in opposite directions and one stationary. The carrier is 2.8 GHz with 10 MHz bandwidth, Friis propagation, isotropic antennas, and proportional-fair scheduling. Handover uses an A3 received-power event, 1.5 dB hysteresis, and a 128 ms trigger interval. Modeled-service traces use actual Radio Resource Control (RRC) signaling; image-service traces use ideal RRC.

The native 3GPP pose/control generator supplies 250 packets per second per UE for 30 seconds. Each packet creates one edge task: 22,500 are submitted and 22,463 delivered per trace. Per-user processing work is 1.2, 1.0, and 0.8 ms at full capacity. Local/remote transport takes 2/20 ms; placement changes buffer new work for 200 ms. Available capacity changes at 8, 18, and 26 seconds. Deadlines are 100, 250, or 500 ms; priorities are 1 or 3. The transfer scenarios update at 12, 20, or 24 seconds, or remain stable. Texts are fixed before model calls, and invalid decisions do not install a new contract.

Let $e$ index a worker. The shared placement heuristic uses per-task processor work $w_u$, a common per-user arrival rate $\lambda>0$, remaining active and queued work $Q_e$, and available processing rate $\mu_e>0$. A joint placement $\mathbf e=(e_u)_u$ assigns worker $e_u$ to user $u$. Its offered work rate at worker $e$ is $\rho_e(\mathbf e)=\sum_{u:e_u=e}\lambda w_u$. With anticipation interval $H=1$ s, define estimated queue and service delay $\widehat q_{u,e}(\mathbf e)$ by
\begin{equation}
\widehat q_{u,e}(\mathbf e)=
\frac{Q_e+w_u+H\max\{0,\rho_e(\mathbf e)-\mu_e\}}{\mu_e}.
\end{equation}
Let $\ell_{u,e}$ be transport delay and $m_{u,e}$ placement-change delay, with $m_{u,e}=0$ for an unchanged placement. Using the interpreted priorities and deadlines from Section~\ref{sec:system}, the scheduler minimizes
\begin{equation}
\sum_u \frac{\widehat p_{u,v}}{\widehat d_{u,v}}
\left(\widehat q_{u,e_u}(\mathbf e)+\ell_{u,e_u}
+\frac{m_{u,e_u}}{\lambda H}\right)
\end{equation}
over permitted joint placements. Parenthesized terms have units of time: the heuristic anticipates one second of overload and amortizes placement changes over one second of arrivals.

\subsection{Real Image Execution}
A fixed, seed-42 sample of 96 IIIT5K test images~\cite{mishra2012scene} is used throughout the application comparison. Each UE sends three images per second for 30 seconds. Original image sizes determine the number of 1,200-byte radio fragments; all fragments must arrive for reconstruction. The two traces reconstruct 234/270 and 235/270 images, with two handovers each. The first UE moves at 15 or 25 m/s, the second in the opposite direction at three quarters of that speed.

The radio uses explicit time-division duplexing (TDD) and a 1 MiB Radio Link Control (RLC) unacknowledged-mode transmit buffer. A preliminary 10 KiB configuration truncated large images. We corrected it and reran the complete fixed matrix; the preliminary results are excluded here.

Contracts start at one second and update at 8, 16, and 24 seconds, with 250/500 ms deadlines. Requests wait for their version for at most the largest supported deadline, then check the interpreted deadline before dispatch. Update windows last 1.5 seconds. Background work arrives at four requests per second at worker A during seconds 9--15 in one trace and at B during seconds 18--24 in the other.

Two logical workers share a host and execute Tesseract~\cite{smith2007overview} on images transferred through Hypertext Transfer Protocol (HTTP). Eight training-image requests calibrate service estimates before evaluation. The scheduler selects the allowed worker with the earliest predicted finish, considering active work and queued work ahead of the request. Expired work is removed before service. Exact text comparison uses Unicode Normalization Form C and trims boundary whitespace, retaining case.

All 3,580 foreground and 384 background OCR executions match service logs by request identity, image digest, worker, and parameters. Maximum request-release lag is 16.06 ms. Raw model responses, contract checks, timestamps, radio identities, and source snapshots are retained. The fixed randomized condition order uses seed 42. Timing repeats share inputs; waiting interventions are post-measurement modeled replays, separate from actual API and OCR observations.

%% file: main.bbl
\begin{thebibliography}{10}
\providecommand{\url}[1]{#1}
\csname url@samestyle\endcsname
\providecommand{\newblock}{\relax}
\providecommand{\bibinfo}[2]{#2}
\providecommand{\BIBentrySTDinterwordspacing}{\spaceskip=0pt\relax}
\providecommand{\BIBentryALTinterwordstretchfactor}{4}
\providecommand{\BIBentryALTinterwordspacing}{\spaceskip=\fontdimen2\font plus
\BIBentryALTinterwordstretchfactor\fontdimen3\font minus
  \fontdimen4\font\relax}
\providecommand{\BIBforeignlanguage}[2]{{%
\expandafter\ifx\csname l@#1\endcsname\relax
\typeout{** WARNING: IEEEtran.bst: No hyphenation pattern has been}%
\typeout{** loaded for the language `#1'. Using the pattern for}%
\typeout{** the default language instead.}%
\else
\language=\csname l@#1\endcsname
\fi
#2}}
\providecommand{\BIBdecl}{\relax}
\BIBdecl

\bibitem{letaief2019roadmap}
K.~B. Letaief, W.~Chen, Y.~Shi, J.~Zhang, and Y.-J.~A. Zhang, ``The roadmap to
  {6G}: {AI} empowered wireless networks,'' \emph{IEEE Communications
  Magazine}, vol.~57, pp. 84--90, Aug. 2019.

\bibitem{itu2023imt}
\BIBentryALTinterwordspacing
{International Telecommunication Union}, ``Framework and overall objectives of
  the future development of {IMT} for 2030 and beyond,'' ITU Radiocommunication
  Sector, Recommendation ITU-R M.2160-0, Nov. 2023. [Online]. Available:
  \url{https://www.itu.int/rec/R-REC-M.2160-0-202311-I}
\BIBentrySTDinterwordspacing

\bibitem{shi2016edge}
W.~Shi, J.~Cao, Q.~Zhang, Y.~Li, and L.~Xu, ``Edge computing: Vision and
  challenges,'' \emph{IEEE Internet of Things Journal}, vol.~3, no.~5, pp.
  637--646, Oct. 2016.

\bibitem{mao2017survey}
Y.~Mao, C.~You, J.~Zhang, K.~Huang, and K.~B. Letaief, ``A survey on mobile
  edge computing: The communication perspective,'' \emph{IEEE Communications
  Surveys \& Tutorials}, vol.~19, no.~4, pp. 2322--2358, 2017.

\bibitem{clemm2022intentbased}
\BIBentryALTinterwordspacing
A.~Clemm, L.~Ciavaglia, L.~Z. Granville, and J.~Tantsura, ``Intent-based
  networking - concepts and definitions,'' RFC Editor, RFC 9315, Oct. 2022.
  [Online]. Available: \url{https://www.rfc-editor.org/rfc/rfc9315}
\BIBentrySTDinterwordspacing

\bibitem{jacobs2021hey}
\BIBentryALTinterwordspacing
A.~S. Jacobs, R.~J. Pfitscher, R.~H. Ribeiro, R.~A. Ferreira, L.~Z. Granville,
  W.~Willinger, and S.~G. Rao, ``Hey, {Lumi}! using natural language for
  intent-based network management,'' in \emph{2021 USENIX Annual Technical
  Conference (USENIX ATC 21)}.\hskip 1em plus 0.5em minus 0.4em\relax USENIX
  Association, Jul. 2021, pp. 625--639. [Online]. Available:
  \url{https://www.usenix.org/conference/atc21/presentation/jacobs}
\BIBentrySTDinterwordspacing

\bibitem{angi2025llnet}
A.~Angi, A.~Sacco, and G.~Marchetto, ``{LLNet}: An intent-driven approach to
  instructing softwarized network devices using a small language model,''
  \emph{IEEE Transactions on Network and Service Management}, vol.~22, pp.
  3403--3418, Aug. 2025.

\bibitem{wang2024netconfeval}
C.~Wang, M.~Scazzariello, A.~Farshin, S.~Ferlin, D.~Kosti{\'c}, and M.~Chiesa,
  ``{NetConfEval}: Can {LLMs} facilitate network configuration?''
  \emph{Proceedings of the ACM on Networking}, vol.~2, no. CoNEXT2, pp. 1--25,
  Jun. 2024.

\bibitem{miyaoka2025chatdriven}
\BIBentryALTinterwordspacing
Y.~Miyaoka, M.~Inoue, K.~Urata, and S.~Harada, ``Chat-driven optimal management
  for virtual network services,'' arXiv preprint arXiv:2512.24614, 2025.
  [Online]. Available: \url{https://arxiv.org/abs/2512.24614}
\BIBentrySTDinterwordspacing

\bibitem{martins2026intentdriven}
\BIBentryALTinterwordspacing
J.~Martins, L.~Mokrushin, M.~Orlic, and A.~K. A, ``Intent-driven {6G} service
  orchestration: Grounded translation, validation, and decomposition,'' arXiv
  preprint arXiv:2606.28348, Jun. 2026. [Online]. Available:
  \url{https://arxiv.org/abs/2606.28348}
\BIBentrySTDinterwordspacing

\bibitem{parra-ullauri2026rolebased}
\BIBentryALTinterwordspacing
J.~{Parra-Ullauri}, T.~A. Khan, D.~McHugh, S.~Kapoor, A.~Duke, A.~Hey, and
  A.~{Corston-Petrie}, ``Role-based agentic {AI} for intent-driven network and
  service orchestration,'' arXiv preprint arXiv:2606.20580, 2026. [Online].
  Available: \url{https://arxiv.org/abs/2606.20580}
\BIBentrySTDinterwordspacing

\bibitem{typesafe2026systemone}
\BIBentryALTinterwordspacing
{TypeSafe AI}, ``Introducing system one models \& {Jev},'' Technical blog,
  2026, accessed September 19, 2026. [Online]. Available:
  \url{https://typesafe.ai/blog/introducing-system-one-models-and-jev}
\BIBentrySTDinterwordspacing

\bibitem{manias2024intentbased}
D.~M. Manias, A.~Chouman, and A.~Shami, ``Towards intent-based network
  management: Large language models for intent extraction in {5G} core
  networks,'' in \emph{2024 20th International Conference on the Design of
  Reliable Communication Networks (DRCN)}.\hskip 1em plus 0.5em minus
  0.4em\relax Montreal, QC, Canada: IEEE, May 2024, pp. 1--6.

\bibitem{lira2025network}
\BIBentryALTinterwordspacing
O.~G. Lira, O.~M. Caicedo, and N.~L.~S. Da~Fonseca, ``Network
  self-configuration based on fine-tuned small language models,'' arXiv
  preprint arXiv:2512.02861, 2025. [Online]. Available:
  \url{https://arxiv.org/abs/2512.02861}
\BIBentrySTDinterwordspacing

\bibitem{wu2024netllm}
D.~Wu, X.~Wang, Y.~Qiao, Z.~Wang, J.~Jiang, S.~Cui, and F.~Wang, ``{NetLLM}:
  Adapting large language models for networking,'' in \emph{Proceedings of the
  ACM SIGCOMM 2024 Conference}.\hskip 1em plus 0.5em minus 0.4em\relax Sydney
  NSW Australia: ACM, Aug. 2024, pp. 661--678.

\bibitem{islam2026intentpreprint}
\BIBentryALTinterwordspacing
K.~Islam and R.~N. Calheiros, ``{Intent Engine}: Natural-language intent
  translation for intent-driven orchestration in the compute continuum,'' arXiv
  preprint arXiv:2608.20388, 2026. [Online]. Available:
  \url{https://arxiv.org/abs/2608.20388}
\BIBentrySTDinterwordspacing

\bibitem{mekrache2026dmogpt}
A.~Mekrache, A.~Ksentini, and C.~Verikoukis, ``{DMO-GPT}: An intent-driven
  framework for distributed {6G} management and orchestration,'' \emph{IEEE
  Communications Magazine}, vol.~64, no.~1, pp. 48--54, Jan. 2026.

\bibitem{brodimas2025intentbased}
D.~Brodimas, A.~Birbas, D.~Kapolos, and S.~Denazis, ``Intent-based
  infrastructure and service orchestration using agentic-{AI},'' \emph{IEEE
  Open Journal of the Communications Society}, vol.~6, pp. 7150--7168, 2025.

\bibitem{li2025jaunt}
\BIBentryALTinterwordspacing
E.~Li and H.~Du, ``{JAUNT}: Joint alignment of user intent and network state
  for {QoE}-centric {LLM} tool routing,'' arXiv preprint arXiv:2510.18550,
  2025. [Online]. Available: \url{https://arxiv.org/abs/2510.18550}
\BIBentrySTDinterwordspacing

\bibitem{nisiotis2026prompt}
\BIBentryALTinterwordspacing
L.~Nisiotis and A.~Hadjiliasi, ``From prompt to service: An {SLM}-based agent
  orchestration gateway for {AI}-driven virtual worlds,'' arXiv preprint
  arXiv:2606.03557, Sep. 2026. [Online]. Available:
  \url{https://arxiv.org/abs/2606.03557}
\BIBentrySTDinterwordspacing

\bibitem{tunstall2022efficient}
\BIBentryALTinterwordspacing
L.~Tunstall, N.~Reimers, U.~E.~S. Jo, L.~Bates, D.~Korat, M.~Wasserblat, and
  O.~Pereg, ``Efficient few-shot learning without prompts,'' arXiv preprint
  arXiv:2209.11055, 2022. [Online]. Available:
  \url{https://arxiv.org/abs/2209.11055}
\BIBentrySTDinterwordspacing

\bibitem{zaratiana2024gliner}
U.~Zaratiana, N.~Tomeh, P.~Holat, and T.~Charnois, ``{GLiNER}: Generalist model
  for named entity recognition using bidirectional transformer,'' in
  \emph{Proceedings of the 2024 Conference of the North American Chapter of the
  Association for Computational Linguistics: Human Language Technologies
  (Volume 1: Long Papers)}.\hskip 1em plus 0.5em minus 0.4em\relax Mexico City,
  Mexico: Association for Computational Linguistics, 2024, pp. 5364--5376.

\bibitem{zaratiana2025gliner2}
U.~Zaratiana, G.~Pasternak, O.~Boyd, G.~{Hurn-Maloney}, and A.~Lewis,
  ``{GLiNER2}: Schema-driven multi-task learning for structured information
  extraction,'' in \emph{Proceedings of the 2025 Conference on Empirical
  Methods in Natural Language Processing: System Demonstrations}.\hskip 1em
  plus 0.5em minus 0.4em\relax Suzhou, China: Association for Computational
  Linguistics, 2025, pp. 130--140.

\bibitem{belcak2025small}
\BIBentryALTinterwordspacing
P.~Belcak, G.~Heinrich, S.~Diao, Y.~Fu, X.~Dong, S.~Muralidharan, Y.~C. Lin,
  and P.~Molchanov, ``Small language models are the future of agentic {AI},''
  arXiv preprint arXiv:2506.02153, 2025. [Online]. Available:
  \url{https://arxiv.org/abs/2506.02153}
\BIBentrySTDinterwordspacing

\bibitem{geng2025jsonschemabench}
\BIBentryALTinterwordspacing
S.~Geng, H.~Cooper, M.~Moskal, S.~Jenkins, J.~Berman, N.~Ranchin, R.~West,
  E.~Horvitz, and H.~Nori, ``{JSONSchemaBench}: A rigorous benchmark of
  structured outputs for language models,'' arXiv preprint arXiv:2501.10868,
  2025. [Online]. Available: \url{https://arxiv.org/abs/2501.10868}
\BIBentrySTDinterwordspacing

\bibitem{zheng2023sglang}
\BIBentryALTinterwordspacing
L.~Zheng, L.~Yin, Z.~Xie, C.~Sun, J.~Huang, C.~H. Yu, S.~Cao, C.~Kozyrakis,
  I.~Stoica, J.~E. Gonzalez, C.~Barrett, and Y.~Sheng, ``{SGLang}: Efficient
  execution of structured language model programs,'' arXiv preprint
  arXiv:2312.07104, 2023. [Online]. Available:
  \url{https://arxiv.org/abs/2312.07104}
\BIBentrySTDinterwordspacing

\bibitem{kwon2023efficient}
W.~Kwon, Z.~Li, S.~Zhuang, Y.~Sheng, L.~Zheng, C.~H. Yu, J.~Gonzalez, H.~Zhang,
  and I.~Stoica, ``Efficient memory management for large language model serving
  with {PagedAttention},'' in \emph{Proceedings of the 29th Symposium on
  Operating Systems Principles}.\hskip 1em plus 0.5em minus 0.4em\relax Koblenz
  Germany: ACM, Oct. 2023, pp. 611--626.

\bibitem{chen2023frugalgptb}
\BIBentryALTinterwordspacing
L.~Chen, M.~Zaharia, and J.~Zou, ``{FrugalGPT}: How to use large language
  models while reducing cost and improving performance,'' arXiv preprint
  arXiv:2305.05176, 2023. [Online]. Available:
  \url{https://arxiv.org/abs/2305.05176}
\BIBentrySTDinterwordspacing

\bibitem{ong2024routellma}
\BIBentryALTinterwordspacing
I.~Ong, A.~Almahairi, V.~Wu, W.-L. Chiang, T.~Wu, J.~E. Gonzalez, M.~W. Kadous,
  and I.~Stoica, ``{RouteLLM}: Learning to route {LLMs} with preference data,''
  arXiv preprint arXiv:2406.18665, 2024. [Online]. Available:
  \url{https://arxiv.org/abs/2406.18665}
\BIBentrySTDinterwordspacing

\bibitem{bang2023gptcache}
F.~Bang, ``{GPTCache}: An open-source semantic cache for {LLM} applications
  enabling faster answers and cost savings,'' in \emph{Proceedings of the 3rd
  Workshop for Natural Language Processing Open Source Software (NLP-OSS
  2023)}.\hskip 1em plus 0.5em minus 0.4em\relax Singapore, Singapore:
  Association for Computational Linguistics, 2023, pp. 212--218.

\bibitem{santos2021lowlatency}
J.~Santos, T.~Wauters, B.~Volckaert, and F.~De~Turck, ``Towards low-latency
  service delivery in a continuum of virtual resources: State-of-the-art and
  research directions,'' \emph{IEEE Communications Surveys \& Tutorials},
  vol.~23, no.~4, pp. 2557--2589, 2021.

\bibitem{crankshaw2020inferline}
D.~Crankshaw, G.-E. Sela, X.~Mo, C.~Zumar, I.~Stoica, J.~Gonzalez, and
  A.~Tumanov, ``{InferLine}: Latency-aware provisioning and scaling for
  prediction serving pipelines,'' in \emph{Proceedings of the 11th ACM
  Symposium on Cloud Computing}.\hskip 1em plus 0.5em minus 0.4em\relax Virtual
  Event USA: ACM, Oct. 2020, pp. 477--491.

\bibitem{patriciello2019nr}
N.~Patriciello, S.~Lagen, B.~Bojovic, and L.~Giupponi, ``An {E2E} simulator for
  {5G NR} networks,'' \emph{Simulation Modelling Practice and Theory}, vol.~96,
  p. 101933, 2019.

\bibitem{koutlia2022calibration}
K.~Koutlia, B.~Bojovic, Z.~Ali, and S.~Lag{\'e}n, ``Calibration of the
  {5G-LENA} system level simulator in {3GPP} reference scenarios,''
  \emph{Simulation Modelling Practice and Theory}, vol. 119, p. 102580, Sep.
  2022.

\bibitem{mishra2012scene}
A.~Mishra, K.~Alahari, and C.~Jawahar, ``Scene text recognition using higher
  order language priors,'' in \emph{Proceedings of the British Machine Vision
  Conference}.\hskip 1em plus 0.5em minus 0.4em\relax Surrey: British Machine
  Vision Association, 2012, pp. 127.1--127.11.

\bibitem{smith2007overview}
R.~Smith, ``An overview of the {Tesseract} {OCR} engine,'' in \emph{Ninth
  International Conference on Document Analysis and Recognition (ICDAR 2007)
  Vol 2}.\hskip 1em plus 0.5em minus 0.4em\relax Curitiba, Parana, Brazil:
  IEEE, Sep. 2007, pp. 629--633.

\end{thebibliography}
